\documentclass[sigconf]{acmart}

\AtBeginDocument{%
  \providecommand\BibTeX{{%
    \normalfont B\kern-0.5em{\scshape i\kern-0.25em b}\kern-0.8em\TeX}}}

\setcopyright{acmcopyright}
\copyrightyear{2023}
\acmYear{2023}
\acmDOI{XXXXXXX.XXXXXXX}

\acmConference[AdKDD '23]{AdKDD '23}{August 07--09, 2023}{Long Beach, CA}
\begin{document}

\title{Towards the  Better Ranking Consistency: A Multi-task Learning Framework for Early Stage Ads Ranking}

\author{Xuewei Wang, Qiang Jin, Shengyu Huang, Min Zhang, Xi Liu, Zhengli Zhao, Yukun Chen, Zhengyu Zhang, Jiyan Yang, Ellie Wen, Sagar Chordia, Wenlin Chen, Qin Huang}
\email{{xwwang, qjin, syhuang,  mzhang27, xliu1, zhengliz, cyk, zhengyuzhang, chocjy, dwen,
sagarc, wenlinchen, huginhuang}@meta.com}
\affiliation{%
  \institution{Meta Platforms, Inc.}
  \streetaddress{1 Hacker Way}
  \city{Menlo Park}
  \state{CA}
  \country{USA}
}






\renewcommand{\shortauthors}{Wang et al.}

\begin{abstract}
 Dividing ads ranking system into retrieval, early, and final stages is a common practice in large scale ads recommendation to balance the efficiency and accuracy. The early stage ranking often uses efficient models to generate candidates out of a set of retrieved ads. The candidates are then fed into a more computationally intensive but accurate final stage ranking system to produce the final ads recommendation. As the early and final stage ranking use different features and model architectures because of system constraints, a serious ranking consistency issue arises where the early stage has a low ads recall, i.e., top ads in the final stage are ranked low in the early stage. In order to pass better ads from the early to the final stage ranking, we propose a multi-task learning framework for early stage ranking to capture multiple final stage ranking components (i.e. ads clicks and ads quality events) and their task relations. With our multi-task learning framework, we can not only achieve serving cost saving from the model consolidation, but also improve the ads recall and ranking consistency. In the online A/B testing, our framework achieves significantly higher click-through rate (CTR), conversion rate (CVR), total value and better ads-quality (e.g. reduced ads cross-out rate) in a large scale industrial ads ranking system.

\end{abstract}



\received[accepted]{21 June 2022}

\maketitle

\section{Introduction}

The goal of the ads ranking system is to select the optimal ads to display to users. Due to latency constraints, it is impractical to predict ranking score for each ad out of large-scale candidates. Therefore, a multi-stage ranking process is widely adopted, which uses progressively more complex models to narrow down the number of ads \cite{covington2016deep,gallagher2019joint,raykar2010designing}. Common multi-stage ranking systems consist of retrieval, early stage ranking, and final stage ranking, as shown in figure~\ref{fig:1}. While retrieval is often rule-based, both early stage and final stage ranking use ranking score predicted by machine learning models.

After we obtain the final stage ads ranking score, the system will run an ads auction to decide the winning set of ads to show to the user. To ensure that the winning ad maximizes value for both user and businesses, we use \textit{total value} \cite{meta} to rank the ads in auction. The total value is a combination of three major factors: 1) The bid placed by an advertiser for that ad. 2) Estimated action rates representing the probability
of the desired outcome (e.g. click, conversion) after showing the ad to a user. 3) Ads quality \cite{quality} capturing the feedback from user on their ads experience.

In general, it is determined by ads quality models, which predict scores of multiple quality events (e.g. crossing out ads, hiding ads). Our framework mainly focuses on learning estimated action rates (i.e. ads CTR) and ads quality.

Despite the multi-stage ads ranking system being a common practice, it has the fundamental problem of multi-stage inconsistency: the early stage ranking system fails to pass good ads to the final stage ranking system \cite{gu2022ranking}. In other words, the low recall of the early stage ranking system can significantly harm the end-to-end ads ranking system. Specifically, we need to resolve the following three challenges in designing effective multi-stage ranking system:

\begin{figure}[t]
  \centering
  \includegraphics[width=0.6\linewidth]{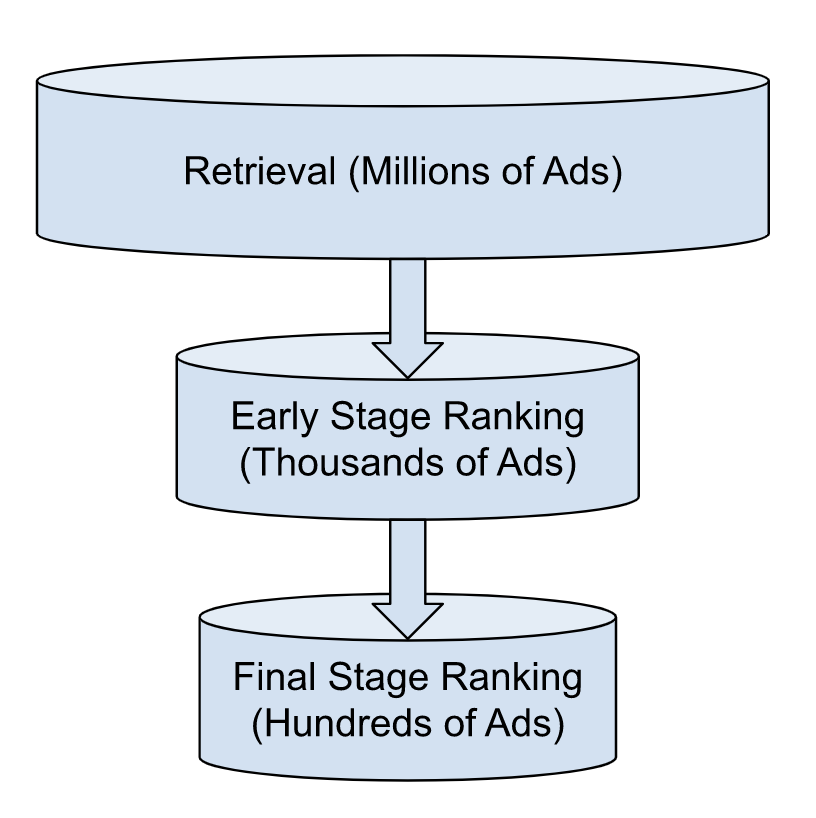}
  \caption{Multi-stage Ranking System Overview}
  \label{fig:1}

\end{figure}

\begin{enumerate}

\item {\textbf{Performance gap between early and final stage}} Due to the restricted model capacity and the smaller feature set, the performance of early stage ads ranking is inferior to that of the final stage ranking. Consequently, when provided with the same candidates, the top ranked ads produced by the final stage ranking and early stage stages can vary a lot.

\item {\textbf{Total value definition inconsistency}} Ideally, we should setup same ranking objectives in the early stage as the final stage, in order to share the same ads total value definition. However, maintaining same types of ads quality models in early stage is difficult considering the heavy engineering work on multiple models and the increased serving cost. To save resource and rank more ads, we only enable major ads quality models in early stage, which causes the ranking consistency issue between the early and final stage ranking.

\item{\textbf{Selection bias}} Conventional early stage ads ranking models are trained on ads with user impression, as well as logged user click or conversion. However, the early stage ads ranking model needs to infer over whole early stage ads candidates, most of which are non-impression ads.
Due to the skewed observed label, the selection bias occur with the distribution mismatch between test and training set \cite{gu2022ranking, chen2023bias,ma2018entire}.
\end{enumerate}

In order to address those issues, we propose a multi-task learning framework for early stage ranking to learn the relevant information of ads total value in final stage ranking. Due to the latency constraint, we cannot learn all components of ads total value in one light-weighted early stage ranking model. Instead, we focus on joint learning of ads CTR and ads quality events. There are three major benefits for our framework:

\begin{itemize}

\item {\textbf{Ranking consistency improvement}}
In order to solve total value definition inconsistency issue on ads quality, we present a new objective for early stage ads quality, called consolidated quality score (CQS). Instead of replicating every final stage ads quality event model in early stage, the CQS consolidates all final stage ads quality objectives together to be a single objective. We derive the CQS task label from the final stage total quality scores. In addition, we add a distillation task from the final stage CTR model. Both of tasks significantly improve the ads recall for early stage ranking.

\item {\textbf{Resource saving by model consolidation}}
In ads auction, the CTR model's prediction is essential to estimate the action rates for various post-click conversions, while ads quality models are necessary for determining the quality score of each ad. Consequently, the primary serving costs stem from the ads CTR model and quality models, due to their large serving traffic. With the multi-task learning for CTR and ads quality events, we can reduce serving costs with shared model architectures and features.

\item {\textbf{Mitigation of selection bias}}
We leverage the data augmentation to mitigate the selection bias of early stage model. We logged more final stage non-impression data in the training data as the augmented data. When computing the CTR loss, instead of treat them as negative samples, we use the final stage CTR prediction as the pseudo-label. For CQS task, the augmented data also has its label as each non-impression ads in final stage still participate in ads auction.

\end{itemize}

In order to better understand the impact of jointly learning CTR and ads quality, we also build a offline recall simulation framework. In the current multi-stage ranking system, the final stage has more accurate prediction for higher precision, whereas the early stage need to optimize for recall. The results show that our framework can improve simulated soft recall for early stage ranking. In the online experiment, we also observe the reduction of total value divergence between early stage and final stage, which implies better ranking consistency. We also conduct ablation study for each key component in our multi-task learning framework. In the online A/B testing, our framework achieves better ads quality, CTR and CVR, compared with the separate serving baseline.

\begin{figure*}
  \centering
  \includegraphics[width=0.7\linewidth]{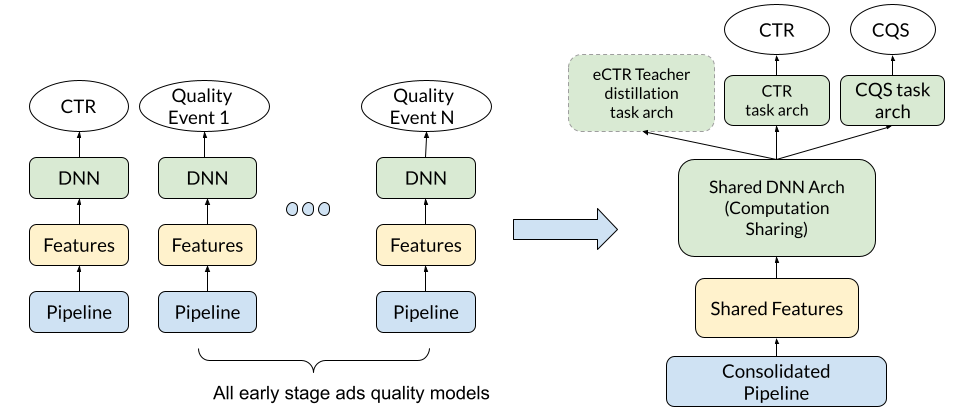}
  \caption{The overview of our multi-task learning framework for early stage ads ranking. We consolidate the ads CTR model and all ads quality models into one multi-task learning model with shared model architecture and features. The CQS denotes for consolidated quality score.}
  \label{fig:mtml}
\end{figure*}

\section{Related Work}
Early stage ads ranking, also known as the pre-ranking stage, has great potential to improve overall ranking performance as they decide candidates for final stage ads ranking. Most of the prior work discussed how to improve the effectiveness while maintain efficiency for early stage ads ranking
\cite{gallagher2019joint,qin2022rankflow,ma2021towards}. Recent work \cite{gu2022ranking} noticed the ranking consistency issue between stages. They introduced a metric, similar to recall, to measure ranking consistency. Also, they conducted experiments for different final stage distillation techniques to improve early stage ranking consistency. However, they only considered improving dedicated ranking models cross stages (e.g. CTR model), but overlooked the interactions between multi-objectives in complex ads ranking system, such as ads quality. Such interaction can be captured through multi-task learning framework.

Multi-task learning is widely used in recommendation system \cite{ma2018modeling,zhao2019recommending,tang2020progressive}. However, prior work mainly focused on complex multi-task learning architectures (e.g. MMoE \cite{ma2018modeling}) to model task relationships. Although those techniques have achieved promising improvements on all tasks, they are difficult to apply in early stage ranking due to model capacity constraint. Recently, a online multi-task framework for CTR and two ads quality models is presented in \cite{ma2022online}. They built a framework which achieved both CTR lift and better ads quality. This framework can not be generalized to different ads ranking ranking systems, which have different ads quality events in the final stage ads ranking. Also, their framework is too complex to use in early stage ranking. Therefore, we still lack simple and efficient work for early stage ranking system to apply multi-task learning. To the best of our knowledge, our work is the first paper discussing the practice for multi-task learning on early stage ranking, from the perspective of ranking consistency and ads CTR-quality joint optimization.

\section{Methods}
In this section, we discuss the key components for our framework: model architecture, model training, and evaluation metrics.

\subsection{Model Architecture}
 Instead training separate early stage CTR and quality models, we propose a multi-task learning framework to train a single model on those objectives, as shown in Figure~\ref{fig:mtml}. We utilize DLRM\cite{DLRM19} framework to build a two-tower model with the user tower and the ad tower. After we obtain the output hidden embeddings from the shared model architecture, we pass them into dedicated task module to learn three tasks. Compared with the original CTR model, we add two additional tasks:

\subsubsection{Consolidated Quality Score (CQS)}
Learning all quality events in a single model can be challenging. First, the data collection process of different quality events varies significantly, which makes it difficult to log all quality events in one data pipeline. For instance, there could be quality events derived from survey-based assessments, whose logging infra that is different from the one used for logging the CTR training data. Moreover, it is challenging for a single model to fulfill the model capacity constraint for efficient inference, while still predicting multiple tasks for quality events.
To address these issues, we propose Consolidated Quality Score (CQS) to consolidate all quality events in early stage ranking. We define the CQS in Equation 2, as the input of the mapping function $f$ to compute the $AdQuality$, which denotes the final ads quality score of an ad. The $pQualityEvent_i$ indicates the model prediction of the quality event $i$. The $scalar_i$ is the associated multiplier, so as to control the quality event's power in the ads auction. The CQS can be easily logged into training data during the ads auction.

\begin{eqnarray}
\label{eq2}
Ad Quality = f(CQS) \\
CQS = \sum_{i=1}^N scalar_i *  pQualityEvent_i
\end{eqnarray}

With the final stage CQS as the label, we not only unblock the quality data logging, but also solve the total value definition inconsistency issue in early stage ranking. Also, the early stage CQS can adapt to final stage quality event changes automatically in a flexible manner and maintain stable multi-stage status. We utilize mean square error as the loss function:
\begin{equation} \label{eq5}
    L_{cqs}= \frac{1}{n} \sum_{i=1}^{n}(CQS_i - y_{cqs})^2,
\end{equation}
where the $CQS_i$ is the final stage consolidated quality score, and $y_{cqs}$ is the early stage ranking predicted value. $n$ is the number of samples.

\subsubsection{CTR Cross-stage Distillation}
In addition to the CTR task and CQS task, we also add one more task for teacher distillation. This task is not used for serving. There are two benefits for using final stage pCTR as the teacher model to distill early stage CTR model. First, distilling knowledge from a teacher model to a student model is a common approach to improve student model's performance without additional capacity cost \cite{hinton2015distilling}. The final stage CTR model is much more complex compared to early stage CTR model, rendering it a reasonable choice to be a teacher model. Second, using the final stage CTR model as the distillation teacher can improve the ranking consistency since the early stage learns the final stage prediction information directly. Although this task can improve ranking consistency, we cannot use this task to replace the original CTR task during serving, because the model cannot learn good calibration without ground-truth click label.
The distillation logistic regression Loss is employed in our teacher task.
\begin{equation} \label{eq3}
L_{teacher} = -[eCTR * \log(y_{ctr}) + (1-eCTR) * \log(1-y_{ctr})],
\end{equation}
where $eCTR$ is the final stage CTR prediction (between 0 to 1) and $y_{ctr}$ is the CTR task head prediction in our multi-task learning framework. The loss function measures the dissimilarity between the early stage CTR prediction and final stage CTR prediction, which helps improve consistency.

\subsection{Model Training}
\subsubsection{Consolidated Data Pipeline}
The serving traffic of CTR model is the subset of that of quality models. For instance, for post-impression conversion types, they do not need the CTR action to complete the conversion, but they still need quality score to rank. Therefore, compared with original CTR pipeline, we add remaining serving traffic for CQS task in the consolidated pipeline.

During model training, the CTR task will only be trained on its serving traffic to avoid unused feedback loop.

\subsubsection{Data Augmentation with Pseudo-label}
In order to mitigate selection bias, we enrich the data with non-impression ads. We randomly subsample the early stage non-impression ads as the augmented data. We treat the final stage CTR prediction as the pseudo-label for those non-impression data, in order to further improve the ranking consistency. During the online training, we have developed data augmentation framework to logging specific model's prediction in the non-impression data.

\subsubsection{Balance Learning for Different Tasks} During offline experiments, we find adding CQS task leads to negative transfer for CTR task. This is expected since
the correlation between CTR and quality score ads is low and ads quality is designed for relevance and integrity. Considering CTR is an important optimized ad event, we tune the weight of the CQS task in the loss to reduce the negative impact of the CTR task. In the final settings of our framework, we adjust the loss weight of CQS to be 1.5, which has the neutral impact on NE of the CTR task. In addition, adding the CTR teacher task can boost the CTR performance significantly. We tune the task weight of CTR teacher to be 2, in order to achieve best performance for CTR.

\subsection{Evaluation of Early Stage Ads Ranking}

There are several common offline evaluation metrics for ads ranking models, such as Area-Under-ROC (AUC) \cite{zhou2018deep,ma2021towards} and normalized entropy (NE) loss \cite{he2014practical}. However, as early stage ranking models aim to improve recall instead of precision, the improvements on user impression data may not generalize to early stage ads, most of which are non-impression data. Furthermore, those offline evaluation metrics only take the individual model's performance into account, but overlook the combined effect of multiple ranking objectives.

Calculating the accurate recall is impractical considering the large-scale ads candidates in early stage. In order to have a better measurement on recall, we leverage the offline simulated recall for multi-objective early stage ranking system. We replay a small traffic with full ad requests in a simulator, a separate ranking flow but copies all components from production flow. Since the simulator will not serve any production traffic, we can relax the timeouts between stages, to ensure that all ads from retrieval stages are ranked. As is shown in Figure~\ref{fig:3}, after obtaining $N$ ads candidates from the retrieval stage, we will pass all ads in a ads request to the simulator and log top $K$ ads in the replay log. Those top $K$ ads will be marked as positive samples and rest of ads in the same ads request will mark as negative samples. We guarantee the production flow and replay flow has the same amount of final ads candidates to reproduce the production flow. After we have the replay log, we can utilize recall metrics to measure model's offline performance, with top $K$ candidates in replay logs as the golden set. There are two types of recall metrics:

\begin{figure}[h]
  \centering
  \includegraphics[width=\linewidth]{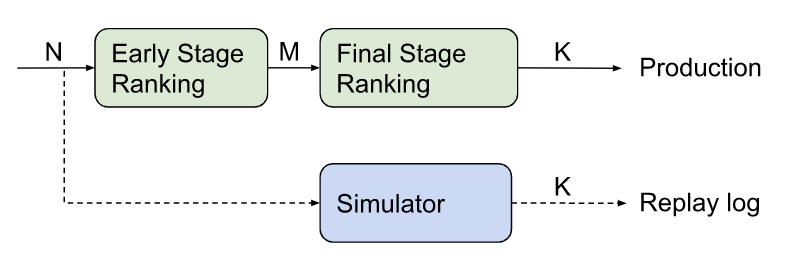}
  \caption{The recall simulator workflow}
  \label{fig:3}

\end{figure}
\begin{itemize}
\item {\textbf{Hard recall}} counts of intersection between top $K$ ads picked by the model and golden set divided by $K$ at the ad request level. This is the widely-accepted definition of recall. The recent work \cite{gu2022ranking} used this as the metrics for ranking consistency.

\item{\textbf{Soft recall}} is the sum of final stage ads total value of top $K$ ads picked by the model divided by sum of total value of the golden set. The hard recall indicates the agreement in terms of ad candidacy, while the soft recall also takes the values of the ads into account.
\end{itemize}

We choose soft recall as the major offline metric for ranking consistency because it is more reasonable for measuring the value of early stage ads. Also, we observe the variance of soft recall among different ad requests is much smaller than hard recall.

\section{Experiments}
In this section, we conduct both offline experiments and online A/B testing to justify the benefit of our framework.
In order to understand each technique better, we first built a simple dedicated CQS model to verify the benefit
of consolidating all early stage quality models. Then we further iterate on the production CTR model with our proposed multi-task learning framework.
For offline metrics, we compare the recall metric for overall multi-task predictions. Compared with other offline metrics, we find the recall metric is more effective to reflect early stage ranking model's online performance, such as impression based total value, CTR, CVR and total value divergence (TVD). The impression based total value is a metric to measure the potential business value of ads after user impression, as we run ads auction depends on the total value of ads. The TVD is computed by the following equation on final stage ads candidates, as an online metric for ranking consistency:

\begin{equation} \label{eq7}
    TVD = \frac{\sum |TotalValue_{final}-TotalValue_{early}|}{\sum |TotalValue_{final}|} \
\end{equation}

For ads quality metrics, we select two quality metrics:

\begin{itemize}
\item \textbf{Ads cross-out (Xout)} happens when a user clicks "\scalebox{0.85}[1]{$\times$}" and selects "I don't want to see this" at the top-right of an ad. We use the ads cross-out rate to measure this quality event, where the lower ads cross-out rate implies better ads quality.

\item{\textbf{Ads Survey for Quality (ASQ)}} is a survey-assessment based metrics for ads quality related signals. It estimates the user rating for ads, where higher is better.
\end{itemize}

\subsection{Consolidate Early Stage Ads Quality Models}
To address the total value definition inconsistency issue between the early and final stages, we study a simple CQS model to consolidate all early stage ads quality models. The offline soft recall shows significant improvement compared with using separate early stage ads quality models. For online metrics, we observe better quality of ads with lower ads cross-out rate and higher ASQ score.
In addition, the total value divergence between early and final stage significantly decreases with the increased impression based total value. Although the CQS model does not affect any CTR or CVR model, the CTR and CVR also increase, which implies the business power of ads quality models. The better ads quality can bring long term value for ads ranking performance, with better ads experience for users. Another benefit for CQS is to save serving CPU cost significantly as we consolidate multiple simple early stage quality models together.
\begin{table}[h]
\begin{tabular}{l|l}
\hline
Recall {\footnotesize (\textbf{+})}   & +3.2\% \\ \hline
Xout rate {\footnotesize (\textbf{-})} &      -1.8\%   \\ \hline
ASQ {\footnotesize (\textbf{+})} & +0.02 \\\hline
TVD {\footnotesize (\textbf{-})} & -7.9\% \\\hline
CTR {\footnotesize (\textbf{+})} & +1.7\% \\\hline
CVR {\footnotesize (\textbf{+})} &  +2.0\% \\ \hline
Total Value {\footnotesize (\textbf{+})} &  +1.0\% \\ \hline
total CPU {\footnotesize (\textbf{-})} & -0.7\% \\\hline
\end{tabular}
\caption{The CQS model's relative performance compared with production early stage quality models. The token {\footnotesize (\textbf{+})} means better performance with higher values, and {\footnotesize (\textbf{-})} means better performance with lower values. }
\label{tab:2}
\end{table}
\subsection{Multi-task Learning of CQS and CTR}
Given the baseline CQS model, we further iterate the CTR model on multi-task framework we proposed. Compared with the production CTR model, we refresh the features add top 50 important CQS features from the CQS model feature importance rank. With more CQS top features, our multi-task learning framework can have neutral MSE performance compared with the baseline CQS model. In table\ref{tab:1}, the multi-learning framework achieves better soft recall than production CTR and baseline CQS models. The online experiment also shows better ads quality and CTR, as well as higher CVR and impression based total value. The total value divergence is further reduced as we add final stage teacher distillation task in our framework. Since we add more features and two more tasks to the original CTR model, the total CPU is slightly smaller than that of separate CTR and CQS models.

\begin{table}[h]
\begin{tabular}{l|l}
\hline

Recall {\footnotesize (\textbf{+})}    & +12.2\% \\ \hline
Xout rate {\footnotesize (\textbf{-})}  &   -3.5\%     \\ \hline
ASQ  {\footnotesize (\textbf{+})} & +0.005 \\\hline
TVD {\footnotesize (\textbf{-})} & -5.7\% \\\hline
CTR {\footnotesize (\textbf{+})} & +0.4\% \\\hline
CVR {\footnotesize (\textbf{+})} & +0.8\% \\\hline
Total Value {\footnotesize (\textbf{+})} &  +3.0\% \\ \hline
total CPU {\footnotesize (\textbf{-})} & -0.06\% \\\hline
\end{tabular}
\caption{The multi-task learning framework's relative performance compared with individual CQS model and CTR model. The token {\footnotesize (\textbf{+})} means better performance with higher values, and {\footnotesize (\textbf{-})} means better performance with lower values. }
\label{tab:1}
\end{table}

\subsection{Ablation Study}
We also set up several comparable models for the ablation study in Table~\ref{tab:3}, in order to exclude the impact of different feature sets compared with production models. We build four baseline models: 1) Dedicated CTR model by removing CQS tasks from our multi-task learning framework. 2) Dedicated CQS model with both CTR and teacher tasks removed. 3) Our multi-task learning framework without teacher task 4)  Our multi-task learning framework trained on impression ads only. According to Figure~\ref{tab:3}, building dedicated CTR and dedicated CQS models can achieve NE or MSE gain over our framework, which implies negative transfer \cite{zhang2022survey} issue in multi-task learning. Without teacher task, the CTR task performance regresses a lot, while the MSE becomes better. The teacher task is essential to help close the performance gap between final stage ranking and early stage ranking.

\begin{table}[h]
\begin{tabular}{l|l|l|l}
\hline
                   & NE diff {\footnotesize (\textbf{-})} & MSE diff {\footnotesize (\textbf{-})} & Recall {\footnotesize (\textbf{+})}\\ \hline
Dedicated CTR + CQS               &    -0.04\% &  -0.6\%  &  -0.6\%  \\
MT w/o Teacher task     & +0.3\%    &  -0.5\%   &  -1.6\%   \\
MT w/o Augmented data   & - &  -  & -11.9\%  \\ \hline
\end{tabular}

\caption{The relative model performance compared with our framework. The MT denotes for our multi-task learning framework. For the model w/o augmented data, the NE and MSE loss are not comparable due to the training data change.}
\label{tab:3}
\end{table}

During the online experiments, the version with dedicated CTR and CQS models shows significant increase on Xout rate and drop for ASQ, although the dedicated CQS model has better MSE performance than our proposed framework. For ads CTR, although the dedicated CTR model can improve the ads CTR with better offline NE performance, the CVR and impression based total value is slightly worse. The higher CTR but lower CVR implies that the ad is very eye-catching, but the user clicking on the ad may not the right demographic for which the ad targets. The poor ads quality can be the explanation of the lower CVR, as the ads quality reflect the user experience on ads. Such results manifest that the single offline metric for a individual ranking model may not be reliable to reflect online performance. The soft recall metric can mitigate this issue which takes multi-objectives into consideration. The larger total value divergence also reflects ranking consistency issue, where the total value between early stage and final stage has large distribution gap. The multi-task learning between CTR and CQS can force the model to learn the coexistence of estimated action rate and ads quality, and improve the total value divergence.

Without the teacher distillation task, the MSE loss for CQS becomes better but the online quality metrics turn out to be worse than our multi-task framework. The online CTR reduces after removing the teacher task, and the total value divergence becomes worse. The impression based total value has regression, which is expected with worse CTR and ads quality.

The augmented data shows great potential in improving early stage ranking performance. After filtering out the augmented data, the offline simulated recall is worst among all baselines. The model also suffers from impression based total value regression, CVR drop and CTR drop, as well as worse ads quality. The augmented data plays a critical role to improve ads recall with selection bias mitigated.

Based on these results, we can draw the following conclusions:

\begin{itemize}
\item Each component in our multi-task learning framework is essential to improve the performance of early stage ads ranking model. The CQS task solves the issue of total value definition inconsistency between early and final stage. The teacher task helps close the performance gap of CTR and improve the ranking consistency. The augmented data mitigates the selection bias with better ads recall.

\item Ads recall and ranking consistency are important for early stage ads ranking. If we only focus on the individual objective of each ads ranking model and optimize for precision, the online overall performance may not improve due to the poor ranking consistency and low ads recall.
\end{itemize}

\begin{table}[t]
\begin{tabular}{l|l|l}
\hline
                    & Xout rate {\footnotesize (\textbf{-})} & ASQ {\footnotesize (\textbf{+})} \\ \hline
Dedicated CTR + CQS      & +7.0\%  & -0.015 \\
MT w/o Teacher task       &  -0.1\%  & -0.002  \\
MT w/o Augmented data    &  +2.8\% &  -0.002  \\ \hline
\end{tabular}

\caption{The relative model online ads quality change compared with our proposed multi-task learning framework.}
\label{tab:4}
\end{table}

\begin{figure}[h]
  \centering
  \includegraphics[width=\linewidth]{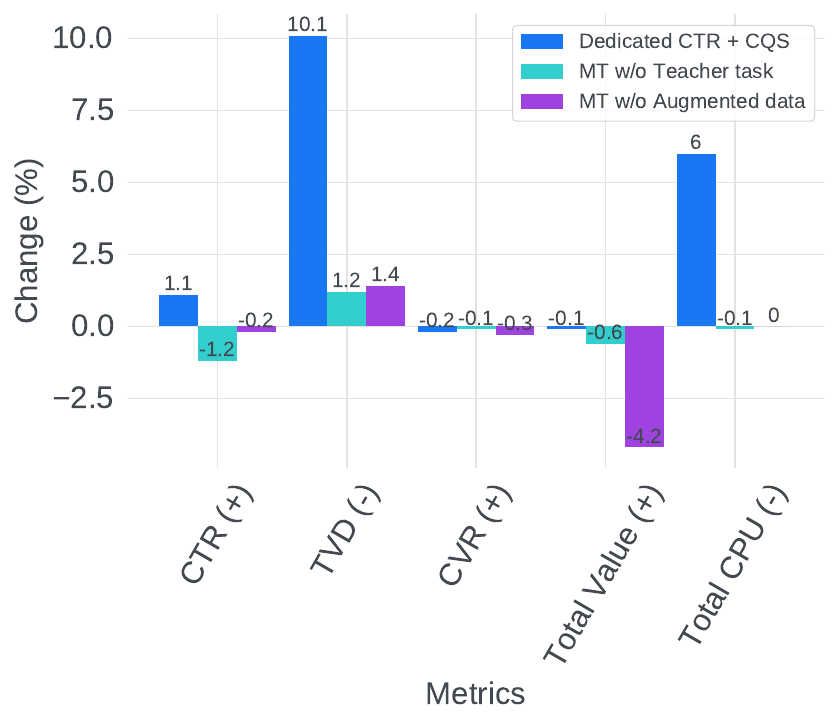}
  \caption{The relative model online performance compared with our proposed multi-task learning framework.}
  \label{fig:1}

\end{figure}

\section{Conclusion and Future work}
We propose a novel multi-task learning framework to improve early stage ads ranking performance. This framework can be generalized to other user cases since the CQS can be applied to any ads ranking system with the ads quality component. We also design the offline recall evaluation metric for the multi-task learning framework in early stage ranking, which has been verified to reflect the model online performance in an industrial ads ranking system.

For future work, we plan to improve the stability of the CQS task. As MSE loss is prone to outliers, we will conduct more experiments for robust regression loss. Also, more techniques  \cite{yu2020gradient,zhang2022survey} can be explored to avoid negative transfer between the CQS and CTR tasks. In addition, we manually tune the weights for different tasks in the current framework. This can be improved with learnable loss weight techniques \cite{kendall2018multi,sener2018multi}, which can adjust weight automatically for multiple tasks.

\bibliographystyle{ACM-Reference-Format}

\end{document}